# Dopant–vacancy complexes as single-photon-emitter candidates in monolayer silicon carbide: radiative and nonradiative decay

Yue Zang[a], Jijun Huang[b], Wenhao Deng[a], Xueling Lei[a *], Song Li[b]

[a]School of Physics, Jiangxi Normal University, Nanchang, Jiangxi 330022, China

[b]Beijing Computational Science Research Center, Beijing, 100193, China

**Abstract:** Solid-state color centers have emerged as promising single-photon sources (SPSs). In particular, color centers in two-dimensional (2D) materials combine high photon brightness with scalable fabrication and ease of integration, offering a versatile platform for quantum-information technologies. Realizing high-efficiency SPSs, however, requires strong radiative emission and negligible nonradiative decay. Using first-principles calculations, we assess radiative and nonradiative decay in dopant–vacancy complexes $X_{Si}V_C$ (X = N, O, P, and S) in monolayer silicon carbide. All four defects exhibit stable neutral charge states and inter-defect optical transitions. Among them, $P_{Si}V_C$ emerges as the leading single-photon-emitter candidate, combining rapid radiative decay with negligible nonradiative decay and a near-unity quantum yield within the modeled channels. By contrast, multiphonon decay limits the quantum yield of $N_{Si}V_C$ despite its high Debye–Waller factor. Hyperfine tensors provide characteristic signatures for identifying the spin-active centers. Beyond identifying promising candidates, these results establish a joint radiative–nonradiative screening strategy and provide theoretical guidance for the rational design of color centers as solid-state SPSs.

[*] Corresponding author. E-mail address: xueling@jxnu.edu.cn

## I. INTRODUCTION

Solid-state color centers have emerged as promising single-photon emitters (SPEs) for quantum-information technologies [1,2], including quantum key distribution [3], quantum repeaters [4], quantum metrology [5], and quantum sensing [6-9]. These defects create atom-like localized states within the host band gap, enabling well-isolated optical transitions and, in many cases, optically addressable spin states, while preserving the robustness and scalability of solid-state hosts.

Diamond color centers established solid-state SPEs [10-13]. However, the $NV^-$ center, despite its excellent spin properties, has a Debye–Waller (DW) factor of only ~3%, with most emission occurring in the phonon sideband. Group-IV centers such as $SiV^-$ and $SnV^-$ offer much higher DW factors (up to 0.7–0.9 for $SiV^-$), yet simultaneously achieving high single-photon efficiency and scalable, cavity-free operation in diamond remains challenging [14-20]. Silicon carbide (SiC) offers a technologically mature alternative to diamond [4,13,21]. Ultrabright, photostable SPSs operating at room temperature have been demonstrated [22-24], and subsequent advances are reviewed in Ref. [25]. However, precise defect positioning and efficient photon extraction remain challenging in bulk crystals. Their high refractive indices limit photon extraction through total internal reflection, while deeply embedded defects are difficult to access and couple to photonic structures. In contrast, 2D hosts place emitters at or near their surfaces, facilitating photon extraction, defect engineering, and heterogeneous integration with photonic and electrical devices [26-28]. Defect-related SPEs in 2D materials have been reported in hexagonal boron nitride [8,29-35], transition-metal dichalcogenides [36-42], and other 2D systems [43,44]. Among these hosts, monolayer SiC is particularly attractive because it combines a wide band gap with nuclear-spin-free elements. Experiments have demonstrated environmentally stable 2D-SiC nanoflakes with visible emission [45]. Nearly planar, millimeter-scale monocrystalline SiC monolayers have also been reported and remain stable in vacuum up to approximately 1200 °C [46]. These results establish the 2D-SiC monolayer as a processable, thermally robust host platform suitable for defect-based quantum emitters. First-principles studies have proposed several native defects and impurity-related centers in monolayer SiC as candidate SPEs. [47-50]. However, the competition between radiative emission and nonradiative decay in these complexes requires systematic assessment to reliably evaluate their potential as high-efficiency SPEs [51,52].

Here, we use first-principles calculations with the HSE06 hybrid functional to investigate dopant–vacancy complexes $X_{Si}V_C$ (X = N, O, P, and S) in monolayer SiC. The $X_CV_{Si}$ color centers have been previously investigated, which have the reverse defect configuration [50]. The assessment covers charge-state stability, localized optical transitions, electron–phonon coupling, and competing radiative and nonradiative decay pathways. Calculated hyperfine tensors and spin densities to characterize the spin-active centers and provide fingerprints for their experimental detection. Among these defects, neutral $P_{Si}V_C$ emerges as the most promising SPE candidate, combining fast radiative decay with negligible nonradiative loss.

## II. COMPUTATIONAL DETAILS

### A. DFT computational parameters

Spin-polarized density functional theory (DFT) calculations were performed using the Vienna *Ab initio* Simulation Package (VASP, version 5.4) [53,54]. Electron–ion interactions were treated using the projector-augmented-wave (PAW) method [55,56], with a plane-wave energy cutoff of 520 eV. Structural relaxations were performed within the Perdew–Burke–Ernzerhof (PBE) generalized-gradient approximation (GGA) [57]. Electronic properties were calculated using the Heyd–Scuseria–Ernzerhof (HSE06) hybrid functional, with an exact exchange fraction of 25% and a screening parameter of 0.2 Å$^{-1}$ [58,59].

Each $X_{Si}V_C$ defect (X = N, O, P, or S) was modeled in an 8 × 8 × 1 supercell of pristine 2D-SiC containing 128 atoms. The supercell was based on our previously reported model [49,50].

Each defect was created by replacing one Si atom with a dopant and removing a neighboring C atom. A vacuum spacing of approximately 20 Å was used to minimize interactions between periodic images. The Brillouin zone was sampled using the Monkhorst–Pack scheme [60], with only the Γ point used for the supercell calculations. Gaussian smearing with a width of 0.05 eV was used. Atomic coordinates were optimized using the conjugate-gradient algorithm while keeping the lattice vectors fixed. The energy and force convergence criteria were set to $10^{-4}$ eV and 0.02 eV $Å^{-1}$, respectively. For electron–phonon coupling calculations, these criteria were tightened to $10^{-5}$ eV and 0.001 eV $Å^{-1}$. Phonon spectra were calculated using PHONOPY [61]. Excited-state geometries were optimized using the constrained-occupation ΔSCF (self-consistent-field) method [62].

### B. Theoretical formalism

The formation energy is defined as

$$E_f^q = E_{tot}^q - E_{tot}^{per} + E_{corr}^q - \sum_i n_i \mu_i + q(E_F + E_{VBM}) \tag{1}$$

where $E_{tot}^q$ and $E_{tot}^{per}$ are the total energies of the defective supercell in charge state $q$ and the neutral pristine supercell, respectively. $E_{corr}^q$ is the finite-size electrostatic correction for the charged 2D supercell. It was evaluated using the Freysoldt–Neugebauer scheme [63] implemented in sxdefectalign2d [64]. Here, $n_i$ is the number of atoms of species $i$ added (positive) to or removed (negative) from the supercell, and $\mu_i$ is the corresponding chemical potential. For the present complexes, $\sum_i n_i \mu_i = \mu_X - \mu_{Si} - \mu_C = \mu_X - (\mu_{Si} + \mu_C) = \mu_X - \mu_{SiC}$, where $\mu_{SiC}$ is the total energy of the 2D-SiC primitive cell. $\mu_X$ is the chemical potential of element X (X = N, O, P, or S) in the reference phase specified in Table S1. The calculated chemical potentials are also reported in Table S1. $E_{VBM}$ denotes the valence-band maximum (VBM) of the pristine supercell. The Fermi level, $E_F$, is referenced to the VBM and varied across the band gap.

The binding energy is defined as

$$E_b(\mathrm{X_{Si}V_C}) = E_f^{min}(\mathrm{X_{Si}}) + E_f^{min}(\mathrm{V_C}) - E_f^{min}(\mathrm{X_{Si}V_C}) \tag{2}$$

where $E_f^{min}(\mathrm{X_{Si}})$ and $E_f^{min}(\mathrm{V_C})$ are the minimum formation energies of the isolated $\mathrm{X_{Si}}$ and $\mathrm{V_C}$ defects, respectively. The quantity $E_f^{min}(\mathrm{X_{Si}V_C})$ is the corresponding minimum formation energy of the defect complex.

The transition dipole moment (TDM) is defined as

$$\mu = \frac{i\hbar}{(E_b - E_a)m}\langle\psi_a|\hat{\mathrm{p}}|\psi_b\rangle \tag{3}$$

where $\psi_a$ and $\psi_b$ are the initial- and final-state wave functions with energy eigenvalues $E_a$ and $E_b$, respectively, $m$ is the electron mass, $\hbar$ is the reduced Planck constant, and $\hat{\mathrm{p}}$ is the momentum operator. The TDMs were calculated at the HSE06 level and processed using the VASP toolkit (vaspkit) [65].

The total Huang–Rhys (HR) factor ($S$) is given by

$$S = \sum_k^M S_k\,,\ S_k = \frac{\omega_k \Delta Q_k^2}{2\hbar} \tag{4}$$

where $M$ is the number of vibrational modes and $k$ is the mode index. The quantities $\omega_k$ and $Q_K$ are the frequency of the $k^{th}$ vibrational mode and the mass-weighted displacement projected onto this mode, respectively.

The photoluminescence (PL) spectrum is given by

$$I_{em}(\hbar\omega) = \omega^3 \cdot \frac{1}{2\pi}\int_{-\infty}^{+\infty} G(t)e^{i\omega t}e^{-\gamma|t|}dt \tag{5}$$

where $\gamma$ is the zero-phonon-line (ZPL) linewidth and $G(t)$ is the generating function. The HR factors and PL spectra were calculated using PyPhotonics [66].

The radiative lifetime is given by [67]:

$$\frac{1}{\tau_{rad}} = \frac{nE_{ZPL}^3|\mu|^2}{3\pi\varepsilon_0 c^3 \hbar^4} \tag{6}$$

where $\tau_{rad}$ is the radiative lifetime, and $n$ is the refractive index of 2D-SiC, taken as approximately 2.65 was used [48]. The quantities $E_{\mathrm{ZPL}}$ and $\mu$ denote the ZPL energy and corresponding TDM, respectively. $\mathcal{E}_0$, $c$ and $\hbar$ are the vacuum permittivity, the speed of light, and the reduced Planck constant, respectively.

Nonradiative transition rates and lifetimes were calculated using Nonrad [68] within a one-dimensional configuration-coordinate model. This approach describes multiphonon transitions between electronic states using Fermi's golden rule in the static-coupling approximation. The nonradiative transition coefficient is given by

$$C = f\frac{2\pi}{\hbar} gV\left|W_{if}\right|^2 \sum_m w_m \sum_n \left|\langle\chi_{im}|\hat{Q} - Q_0|\chi_{fn}\rangle\right|^2 \times \delta(\Delta\varepsilon + m\hbar\Omega_i - n\hbar\Omega_f) \tag{7}$$

where $f$ is a scaling factor, $g$ is the configurational degeneracy, and $V$ is the supercell volume. The configurational degeneracy was set to 1 for all transitions considered here. $W_{if}$ is the electron–phonon coupling matrix element, and $w_m$ is the thermal occupation of the $m$-th vibrational level of the initial state. The initial and final electronic states correspond to the localized excited state and ground state, respectively. These states have vibrational wave functions $\chi_{im}$ and $\chi_{fn}$, respectively, and an energy difference of $\Delta\varepsilon$. Their harmonic phonon frequencies are $\Omega_i$ and $\Omega_f$, respectively. Here, $Q$ is the mass-weighted configuration coordinate, and $Q_0$ is the reference atomic configuration. The mass-weighted displacement between the relaxed initial- and final-state structures is denoted by ΔQ. The values of ΔQ and Δε used in the calculations are listed in Table S2. For these localized-state transitions, $V$ and $f$ were set to unity, so Eq. (7) yields a transition rate in $\mathrm{s}^{-1}$.

The hyperfine tensor ($A_{ij}$) provides a microscopic spin signature and comprises isotropic Fermi-contact and anisotropic dipolar contributions:

$$A_{ij} = \frac{2}{3}\frac{\mu_0\gamma_e\gamma_I\delta_{ij}}{\langle S_z\rangle}\int \delta_T(\vec{r})\rho_s(\vec{r}+\overrightarrow{R_I})d\vec{r} + \frac{\mu_0}{4\pi}\frac{\gamma_e\gamma_I}{\langle S_z\rangle}\int \frac{\rho_s(\vec{r}+\overrightarrow{R_I})}{r^3}\frac{3r_ir_j-\delta_{ij}r^2}{r^2}d\vec{r} \tag{8}$$

where $\mu_0$ is the vacuum permeability, and $\gamma_e$ and $\gamma_I$ are the electron and nuclear gyromagnetic ratios, respectively. Here, $\langle S_z\rangle$ is the expectation value of the z-component of the total electron spin, $\rho_s$ is the spin density, and $\delta_T(\vec{r})$ is a broadening function. The nuclear position is $\overrightarrow{R_I}$, and $r_i$ denotes the $i^{th}$ Cartesian component of $\vec{r}$, with $r = |\vec{r}|$.

## III. RESULTS AND DISCUSSION

### A. Structural reconstruction and charge-state stability

Fig. S1 shows the optimized $X_{Si}V_C$ structures (X = N, O, P, and S) in monolayer SiC. Fig. 1(a) highlights the local reconstruction, with the arrow indicating an enlarged view of the relaxed defect. Two Si atoms adjacent to the C vacancy formed a Si–Si bond, producing a five-membered ring while preserving the $C_{2v}$ symmetry of the defect complex. This reconstruction is common to all four complexes and defines the local environment for the optical transitions discussed below.

The HSE06 density of states (DOS) in Fig. 1(b) shows localized states within the host band gap. $O_{Si}V_C$ and $S_{Si}V_C$ have closed-shell (S=0) ground states, whereas $N_{Si}V_C$ and $P_{Si}V_C$ have doublet ground states (S=1/2). The spin-resolved defect levels and associated optical transitions are examined in Sec. III B.

Fig. 1(c) compares the formation energies of the +1, 0, and -1 charge states, and Table 1 lists the charge-transition levels. $N_{Si}V_C$ changed from positive to neutral at a (+|0) level of 1.190 eV. $O_{Si}V_C$ supported three charge states, with (+|0) and (0|-) transition levels at 0.675 and 2.799 eV, respectively. $P_{Si}V_C$ remained neutral throughout the pristine-supercell band gap, giving it the

broadest neutral-state stability range among the four defects. For $S_{Si}V_C$, the (+|0) and (0|-) levels were 0.709 and 3.323 eV, respectively. Its negative charge state was stable only in the 0.203 eV interval below the conduction-band minimum (CBM), located at 3.526 eV. All four defects possess thermodynamically stable neutral charge states, which are the focus of the following analysis. The positive binding energies in Fig. S2 further confirm that $X_{Si}V_C$ complex formation is thermodynamically favorable relative to isolated $X_{Si}$ and $V_C$ defects.

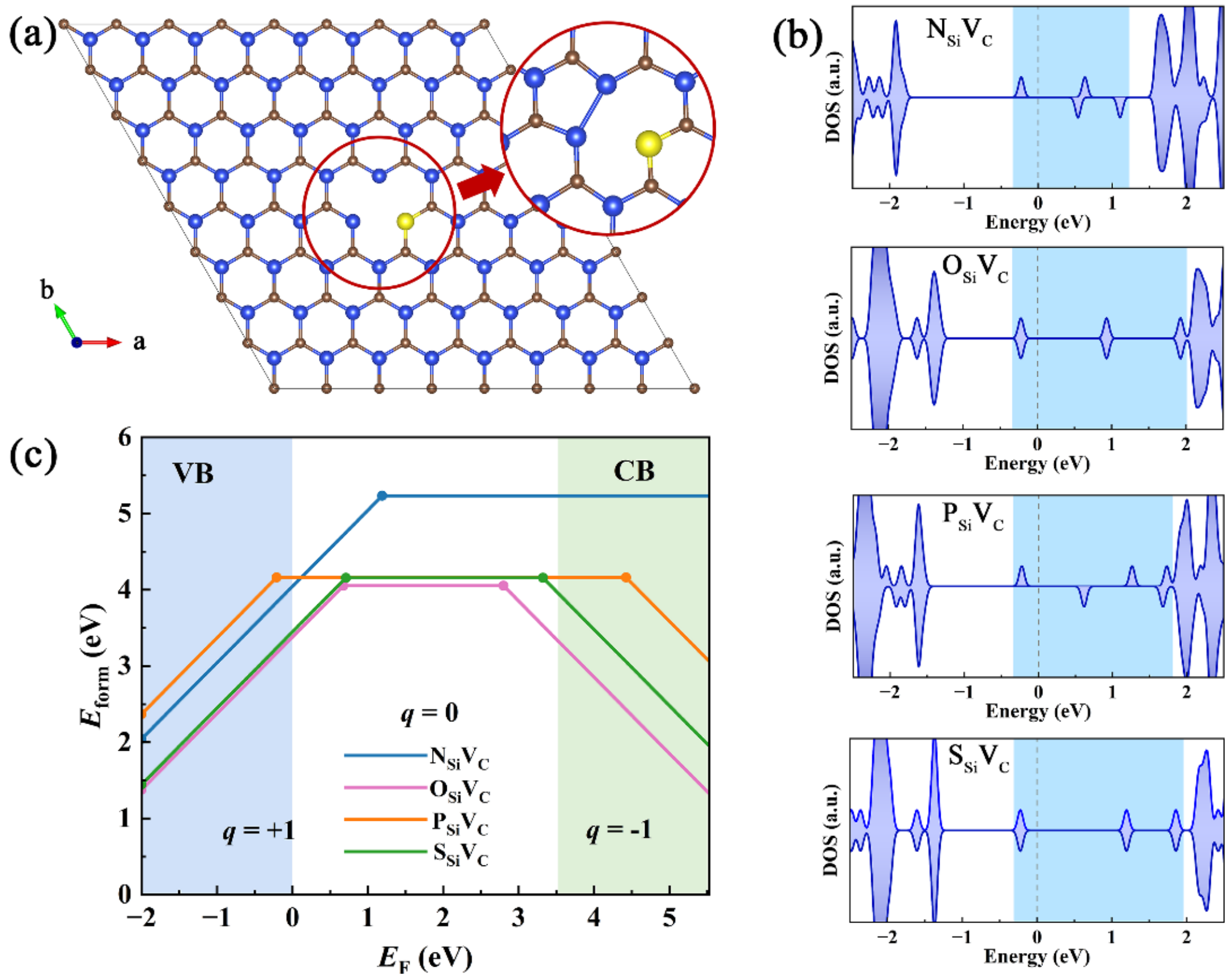

Fig. 1. Defect configurations (a), electronic densities of states (b), and formation energies (c) of $X_{Si}V_C$ (X = N, O, P, and S). Brown, blue, and yellow spheres represent C, Si, and substitutional dopant atoms, respectively. In panel (c), pale blue and green shaded regions indicate the valence band (VB) and conduction band (CB), respectively. Electronic structures and formation energies were calculated using HSE06.

Table 1. HSE06 charge-transition levels ($\varepsilon$) of $X_{Si}V_C$ (X = N, O, P, and S) relative to the pristine-supercell VBM, in eV. The pristine-supercell band gap is 3.526 eV.

| Defects | $\varepsilon(+\|0)$ | $\varepsilon(0\|-)$ |
|---|---|---|
| $N_{Si}V_C$ | 1.190 | 6.658 |
| $O_{Si}V_C$ | 0.675 | 2.799 |
| $P_{Si}V_C$ | -0.209 | 4.425 |
| $S_{Si}V_C$ | 0.709 | 3.323 |

### B. Localized defect levels and optical transitions

Fig. 2 shows the in-gap defect levels and corresponding Kohn–Sham wave functions of neutral $X_{Si}V_C$ defects (X = N, O, P, and S). The levels are labeled by the irreducible representations of the defect point group. The occupied $1b_2$ and lowest unoccupied $2b_2$ states in the spin-up channel supported spin-conserving $1b_2 \rightarrow 2b_2$ transitions in $N_{Si}V_C$ and $P_{Si}V_C$. The occupied $1b_2$ and lowest unoccupied $2b_2$ states in both spin channels supported spin-conserving $1b_2 \rightarrow 2b_2$ transitions in $O_{Si}V_C$ and $S_{Si}V_C$. The wave functions were concentrated around the vacancy, dopant, and neighboring Si–C framework, confirming strongly localized optical excitation.

Table 2 reports the separations of the lowest unoccupied and occupied transition levels from the CBM ($\Delta E_C$) and VBM ($\Delta E_V$), respectively. It also lists their energy difference ($\Delta E$) and the calculated band gap for each defect. The occupied levels lay 1.095–1.593 eV above the VBM, whereas the lowest unoccupied levels lay 0.625–1.179 eV below the CBM. Their separation from the band edges favors localized excitation. The $\Delta E$ values ranged from 0.859 eV for $N_{Si}V_C$ to 1.489 eV for $P_{Si}V_C$. These single-particle energy differences characterize the electronic transitions, whereas emission performance also depends on the relaxed ZPL energies and competing decay rates.

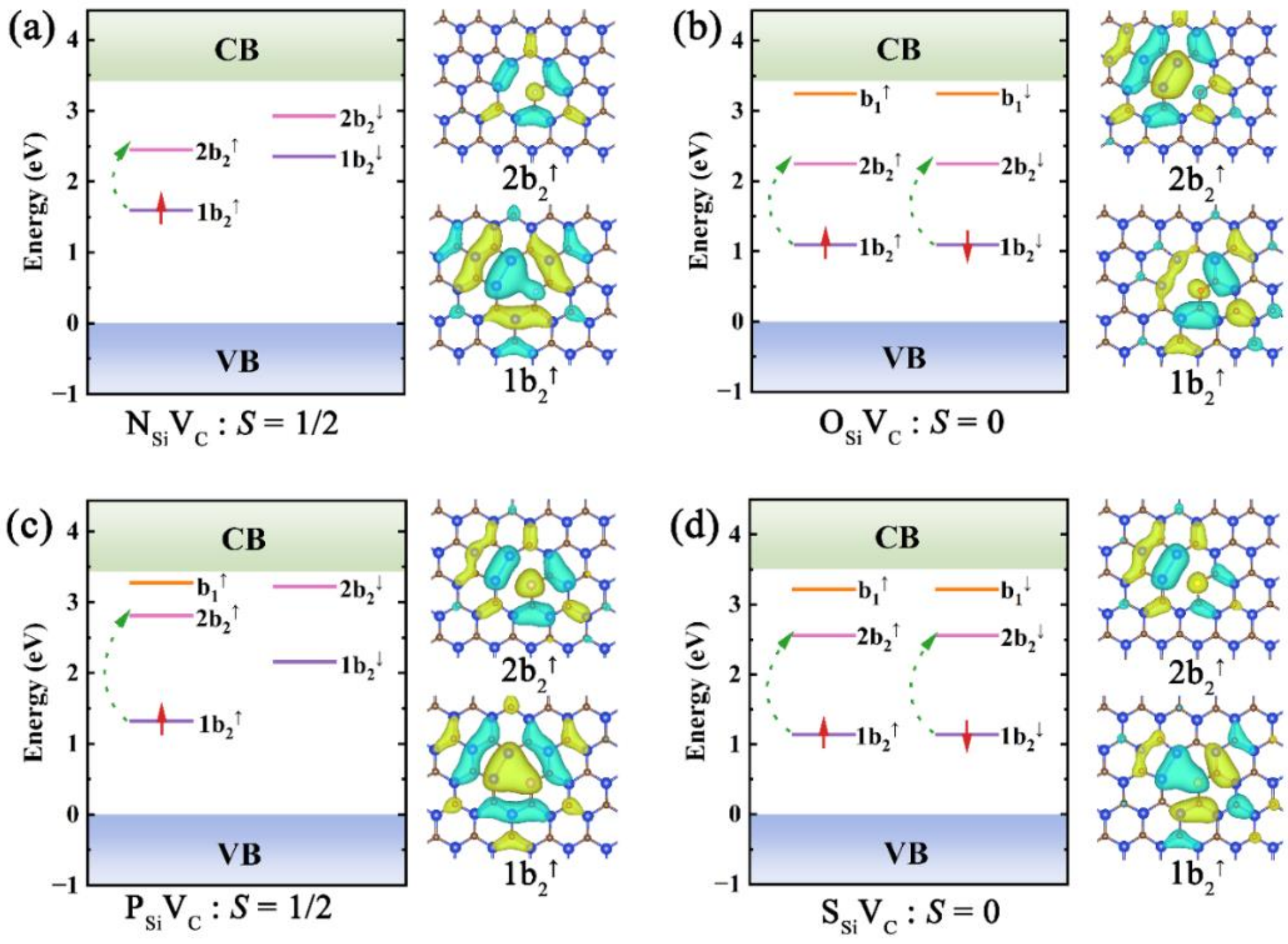

Fig. 2. In-gap defect levels and Kohn–Sham wave functions of neutral $N_{Si}V_C$ (a), $O_{Si}V_C$ (b), $P_{Si}V_C$ (c), and $S_{Si}V_C$ (d), calculated using HSE06. Red arrows denote electron spins, and green arrows denote optical transitions. The defect levels are labeled by their orbital irreducible representations. Pale blue and green shaded regions indicate the VB and CB, respectively. Light-blue and yellow lobes represent negative and positive wave-function isosurfaces, respectively, with an absolute isosurface value of $10^{-9}$ Å$^{-3}$.

Table 2. HSE06 energy separations of the unoccupied transition level from the CBM ($\Delta E_C$) and of the occupied transition level from the VBM ($\Delta E_V$), and the energy difference between these levels ($\Delta E$). All energies are in eV.

| Defect | $\Delta E_C$ | $\Delta E_V$ | $\Delta E$ |
|---|---|---|---|
| $N_{Si}V_C$ | 0.961 | 1.593 | 0.859 |
| $O_{Si}V_C$ | 1.179 | 1.095 | 1.154 |
| $P_{Si}V_C$ | 0.625 | 1.317 | 1.489 |
| $S_{Si}V_C$ | 0.946 | 1.138 | 1.420 |

### C. Magneto-optical properties and spin signatures

An ideal SPE combines high photon purity, indistinguishability, brightness, and stability with a suitable emission wavelength and ease of photonic integration [26]. Here, we assess ZPL energies, TDMs, radiative lifetimes, electron–phonon coupling, and competing relaxation pathways. A high transition energy and large TDM favor rapid radiative decay, whereas the observable photon rate also depends on nonradiative loss and photon extraction. The DW factor quantifies the fraction of emitted light in the ZPL and is distinct from the total luminescence quantum yield.

Table 3 summarizes the optical parameters of the four neutral $X_{Si}V_C$ centers. Their $1b_2^{\uparrow} \rightarrow 2b_2^{\uparrow}$ transitions were optically active because of the large wave-function overlap, with TDMs of 5.4–9.2 D and near-infrared ZPL energies of 0.62–1.15 eV. $P_{Si}V_C$ combined the largest TDM (9.2 D) with the highest ZPL energy (1.15 eV), yielding the shortest radiative lifetime (18.2 ns). $S_{Si}V_C$ had the second-shortest radiative lifetime (33.5 ns). Although $N_{Si}V_C$ had a similarly large TDM (9.0 D), its lower ZPL energy (0.62 eV) yielded a radiative lifetime of 120.4 ns. $O_{Si}V_C$ had the smallest TDM and the longest radiative lifetime. These results show that transition energy and dipole strength must be considered together when assessing radiative emission.

Electron–phonon coupling redistributes emission between the ZPL and phonon sideband (PSB), as illustrated by the PL spectra in Fig. 3. $N_{Si}V_C$, $P_{Si}V_C$, $S_{Si}V_C$, and $O_{Si}V_C$ had HR factors of 0.65, 1.33, 2.01, and 2.53, respectively, corresponding to DW factors of 52%, 26%, 13%, and 8%. $N_{Si}V_C$ exhibited the largest ZPL fraction and the weakest phonon sidebands, whereas the other defects showed progressively stronger phonon-assisted emission. Its DW factor exceeds earlier predictions for other defects in monolayer SiC [47-50]. $P_{Si}V_C$ and $S_{Si}V_C$ also exhibited higher ZPL fractions and shorter radiative lifetimes than $O_{Si}V_C$. We therefore excluded $O_{Si}V_C$ from the subsequent relaxation analysis because it had the longest radiative lifetime (185.3 ns) and lowest DW factor (8%). The total emission efficiency of the remaining candidates also depends on nonradiative relaxation.

Table 3. Optical transitions, TDMs, ZPL energies, HR and DW factors, and radiative lifetimes ($\tau_{rad}$) of neutral $X_{Si}V_C$ defects (X = N, O, P, and S). All calculations were performed at the HSE06 level.

| Defect | Transition path | TDM (D) | ZPL (eV) | HR | DW | $\tau_{rad}$ (ns) |
|---|---|---|---|---|---|---|
| $N_{Si}V_C$ | $1b_2^{\uparrow} \rightarrow 2b_2^{\uparrow}$ | 9.0 | 0.62 | 0.65 | 52% | 120.4 |
| $O_{Si}V_C$ | $1b_2^{\uparrow} \rightarrow 2b_2^{\uparrow}$ | 5.4 | 0.75 | 2.53 | 8% | 185.3 |
| $P_{Si}V_C$ | $1b_2^{\uparrow} \rightarrow 2b_2^{\uparrow}$ | 9.2 | 1.15 | 1.33 | 26% | 18.2 |
| $S_{Si}V_C$ | $1b_2^{\uparrow} \rightarrow 2b_2^{\uparrow}$ | 8.1 | 1.01 | 2.01 | 13% | 33.5 |

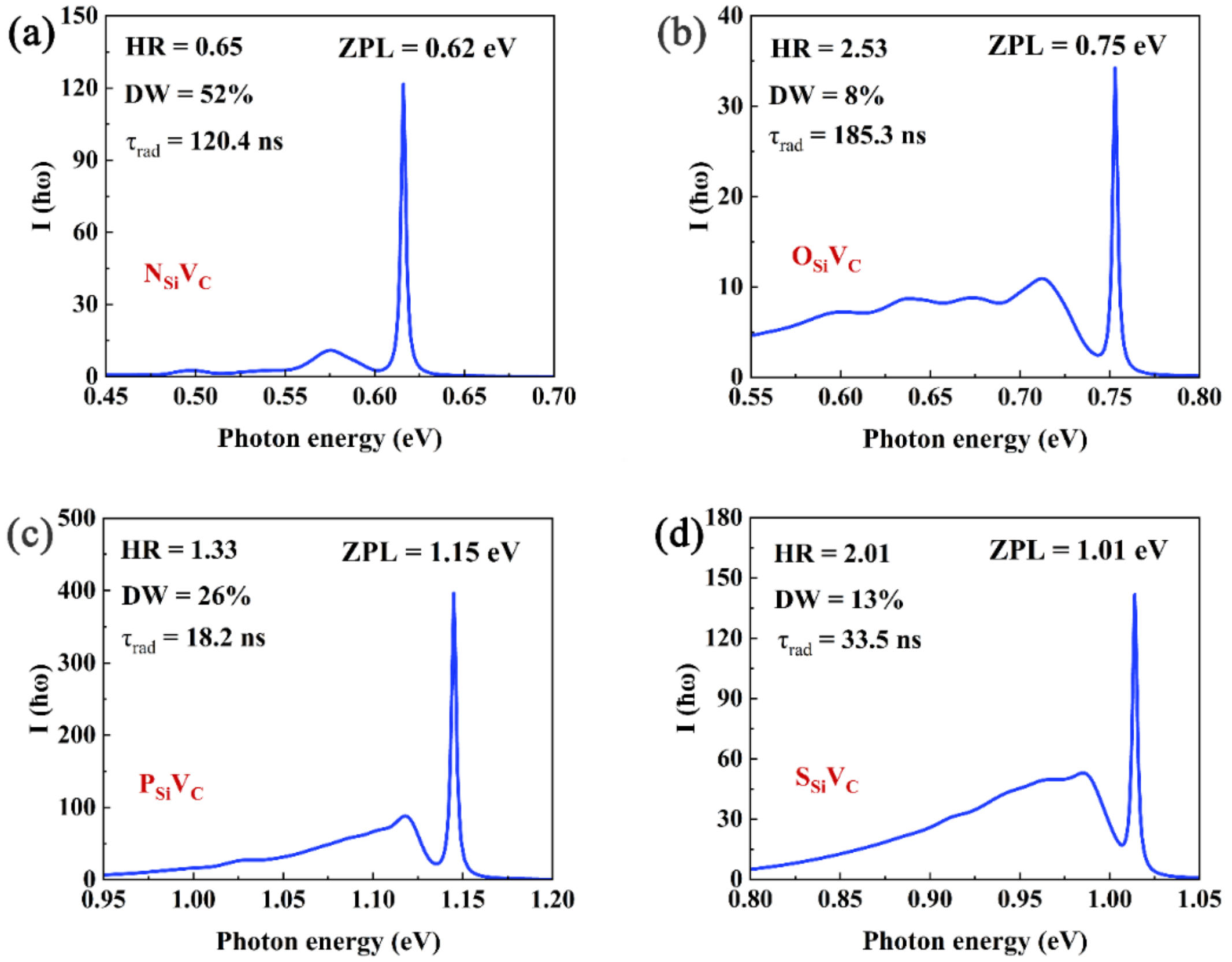

Fig. 3. HSE06 PL spectra for the spin-conserving optical transitions in neutral $N_{Si}V_C$ (a), $O_{Si}V_C$ (b), $P_{Si}V_C$ (c), and $S_{Si}V_C$ (d).

Excited-state relaxation can proceed through fluorescence, spin-conserving internal conversion (IC), or intersystem crossing (ISC) between states with different spin multiplicities. We compared the excited-state energies and spin multiplicities of $N_{Si}V_C$, $P_{Si}V_C$, and $S_{Si}V_C$ to identify possible competing pathways (Table S3). Fig. S3 shows the orbital occupancies, total spins, and energies relative to $S_1$ for the excited states ($S_n$). All examined excited states of $N_{Si}V_C$ and $P_{Si}V_C$ had doublet character, and no lower-energy excited state was identified. For $S_{Si}V_C$, a triplet excited state lay 0.145 eV below the optically excited singlet, providing a possible ISC channel. Population in this triplet could subsequently relax through phosphorescence or nonradiative decay.

Fig. 4 presents one-dimensional configuration-coordinate diagrams, electron–phonon coupling, and temperature-dependent nonradiative lifetimes for $N_{Si}V_C$, $P_{Si}V_C$, and $S_{Si}V_C$.

Panels (a)–(c) show the excited- and ground-state potential-energy surfaces along $Q$, from which the effective phonon energies $\hbar\Omega_i$ and $\hbar\Omega_f$ were obtained. Linear fits to the signed wave-function overlaps versus $Q$ in panels (d)–(f) yield $W_{if}$. The parameters $\hbar\Omega_i$, $\hbar\Omega_f$, and $W_{if}$ enter the expression for the IC-rate in Eq. (7). The reciprocals of the IC rates give the nonradiative lifetimes, shown in panels (g)–(i). These nonradiative lifetimes decreased with increasing temperature, but varied only weakly at low temperatures. For $N_{Si}V_C$, even the longest nonradiative lifetime, 82.2 ns, was shorter than its radiative lifetime of 120.4 ns. This competition limited its predicted luminescence quantum yield to at most approximately 41%, despite a DW factor of 52%. The HR factor characterizes lattice displacement during the optical transition, whereas IC also depends on electronic coupling and vibrational-state overlap. Weak phonon sidebands therefore do not necessarily imply weak nonradiative decay. For $P_{Si}V_C$, the nonradiative lifetime remained much longer than the radiative lifetime up to 800 K. This behavior supports a near-unity quantum yield within the model that includes radiative decay and IC. $S_{Si}V_C$ likewise exhibited negligible IC relative to radiative decay, but its lower-energy triplet provides a possible ISC pathway that could reduce its total quantum yield. This possible additional loss channel makes $P_{Si}V_C$ the leading candidate among the investigated defects.

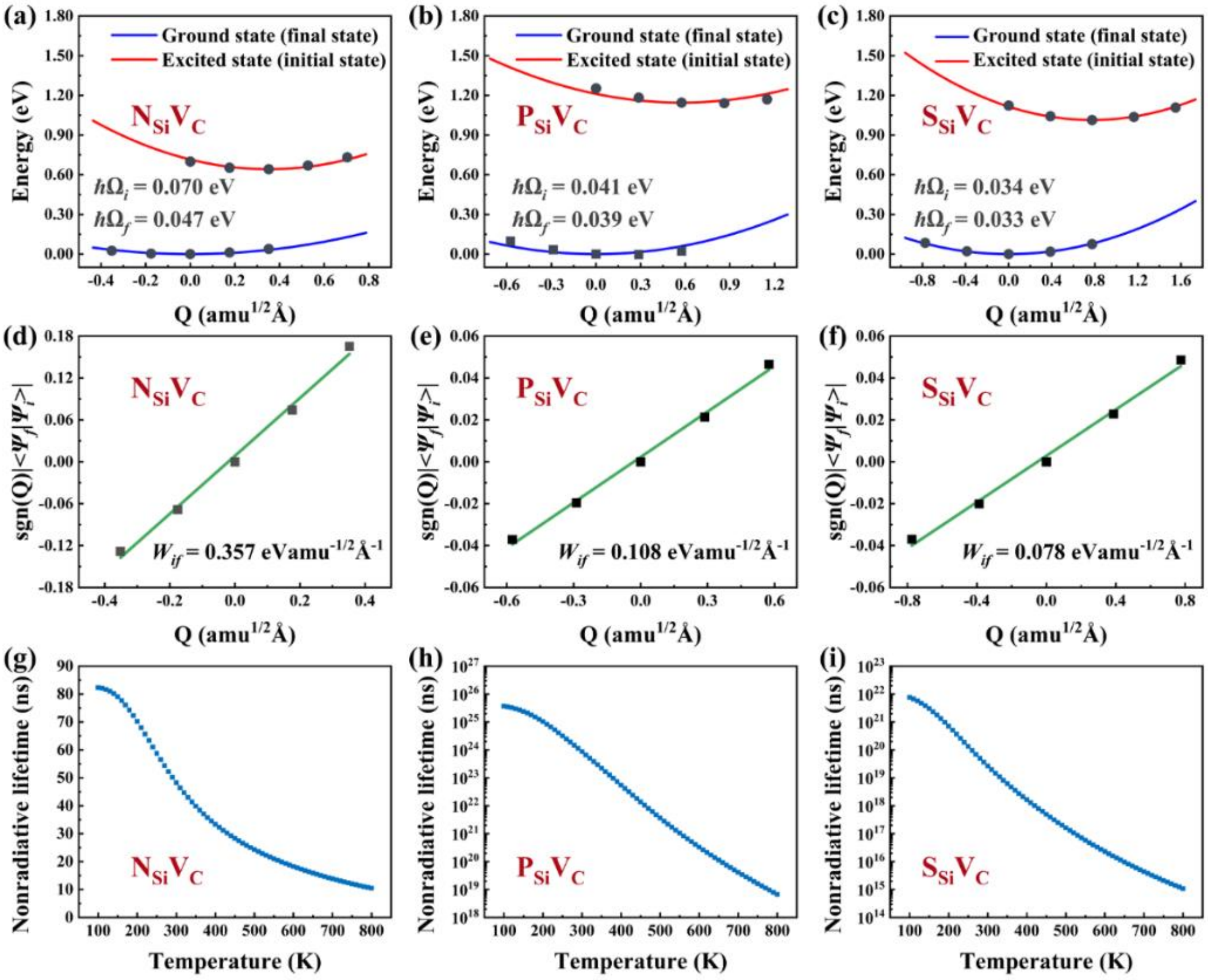

Fig. 4. One-dimensional configuration-coordinate diagrams (a–c), electron–phonon coupling (d–f), and temperature-dependent nonradiative lifetimes (g–i) for the $N_{Si}V_C$, $P_{Si}V_C$, and $S_{Si}V_C$ defects. Panels (a)–(c) show the ground-state (final) and excited-state (initial) potential-energy surfaces as functions of the mass-weighted configuration coordinate $Q$. Symbols denote the calculated energies, and solid curves are parabolic fits used to determine the effective phonon energies $\hbar\Omega_i$ and $\hbar\Omega_f$. Panels (d)–(f) show the signed wave-function overlaps as functions of $Q$. Symbols denote the calculated overlaps, and solid lines are linear fits used to extract the electron–phonon coupling matrix elements $W_{if}$. All calculations were performed at the HSE06 level.

To aid microscopic identification of the four defects, we calculated hyperfine tensors and spin-density distributions. $N_{Si}V_C$ and $P_{Si}V_C$ exhibit strongly anisotropic coupling, as summarized in Table 4. For $N_{Si}V_C$, the N hyperfine principal values are (4.2, 4.2, 19.8) MHz. The strongest Si-site coupling occurs at $Si_3$ and $Si_4$, with $A_{zz}$ = -87.9 MHz, whereas the C-site couplings are comparatively small. $P_{Si}V_C$ exhibits stronger on-site coupling at P, with ($A_{xx}$, $A_{yy}$, $A_{zz}$) = (-60.4, -53.2, 121.9) MHz, alongside appreciable coupling to neighboring Si atoms. These distinct dopant tensors provide microscopic signatures for distinguishing the two spin-active defects.

The positive dopant $A_{zz}$ values and negative values at selected Si sites reflect the combined Fermi-contact and dipolar contributions. These site-resolved hyperfine patterns can help identify constituent atoms and distinguish microscopic models of SiC color centers [4,25]. $O_{Si}V_C$ and $S_{Si}V_C$ have spin-unpolarized ($S$ = 0) ground states, so their ground-state hyperfine tensors vanish within the present approximation.

Table 4. HSE06 hyperfine principal values ($A_{xx}$, $A_{yy}$, and $A_{zz}$) for $^{29}$Si, $^{13}$C, $^{14}$N, and $^{31}$P nuclei at the labeled sites in $N_{Si}V_C$ and $P_{Si}V_C$. Atom labels are defined in Fig. 5.

| Defect | Atoms | $A_{xx}$ (MHz) | $A_{yy}$ (MHz) | $A_{zz}$ (MHz) |
|---|---|---|---|---|
| $N_{Si}V_C$ | N | 4.2 | 4.2 | 19.8 |
| | $C_1$, $C_2$ | -11.3 | -9.5 | -13.1 |
| | $C_3$, $C_4$ | -9.1 | -8.5 | -13.8 |
| | $C_5$, $C_6$ | -5.1 | -2.7 | -5.6 |
| | $C_7$ | 0.9 | 0.7 | 3.4 |
| | $Si_1$, $Si_2$ | -11.9 | -11.8 | -41.8 |
| | $Si_3$, $Si_4$ | -10.0 | -5.5 | -87.9 |
| | $Si_5$ | -5.1 | -5.1 | -26.3 |
| $P_{Si}V_C$ | P | -60.4 | -53.2 | 121.9 |

| | | | |
|---|---|---|---|
| $C_1$, $C_2$ | -1.8 | 0.9 | -3.0 |
| $C_3$, $C_4$ | -6.3 | -5.7 | -6.6 |
| $C_5$, $C_6$ | -3.4 | 0.5 | -3.8 |
| $C_7$ | 1.1 | 0.9 | 3.4 |
| $Si_1$, $Si_2$ | -5.5 | -5.2 | -25.5 |
| $Si_3$, $Si_4$ | -5.8 | -1.4 | -65.4 |
| $Si_5$ | -2.5 | -2.4 | -17.9 |

Fig. 5 shows that the unpaired spin in $N_{Si}V_C$ and $P_{Si}V_C$ is localized within the reconstructed defect complex. In both defects, the spin density extends over the dopant and neighboring Si–C framework. This localization is consistent with the defect-state wave functions in Fig. 2 and the site-resolved hyperfine tensors. Together, the spin-density distributions and tensor anisotropies provide complementary microscopic signatures for identification by electron paramagnetic resonance (EPR).

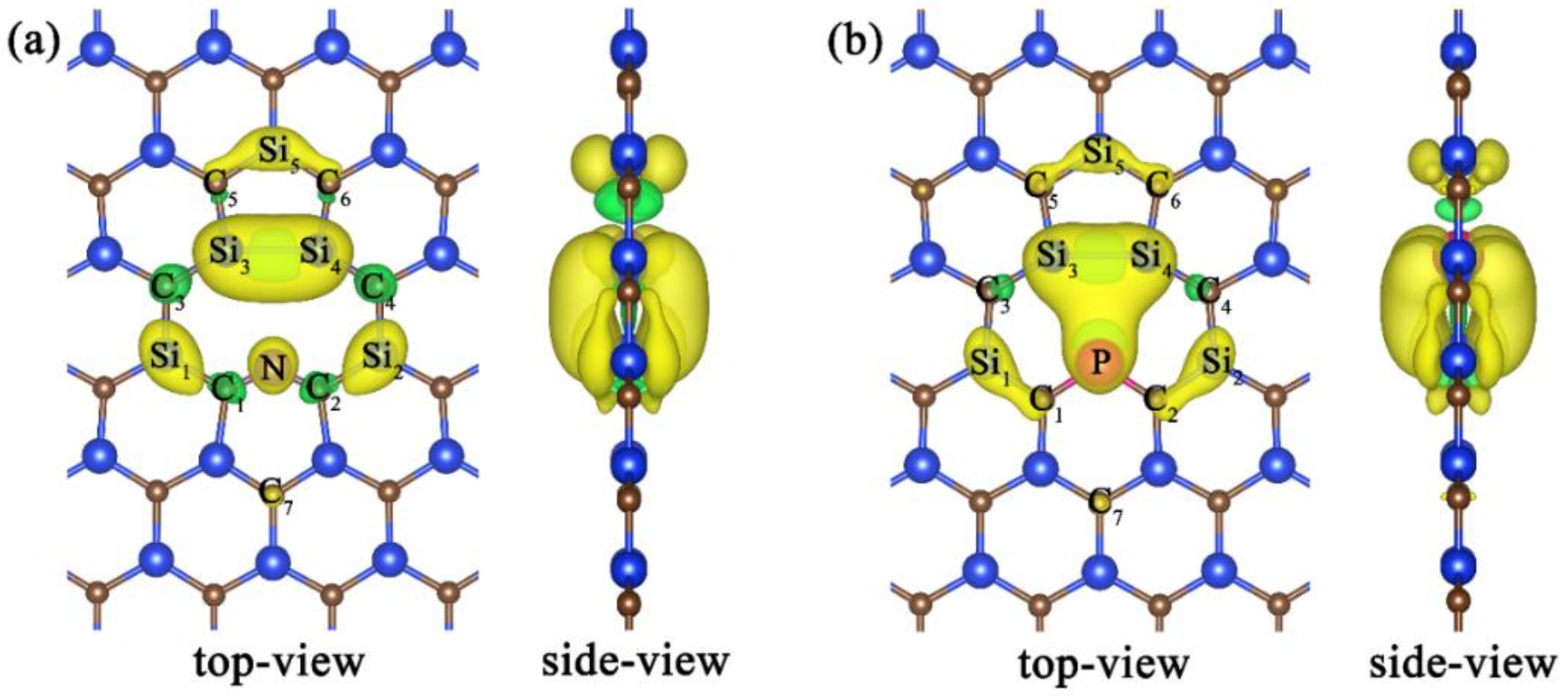

Fig. 5. HSE06 spin-density distributions of $N_{Si}V_C$ (a) and $P_{Si}V_C$ (b) in top and side views. Blue and brown spheres represent Si and C atoms, respectively, while purple and red spheres denote N and P, respectively. Hyperfine constants for the labeled atoms are given in Table 4.

Light-green and yellow lobes represent negative and positive spin-density isosurfaces, respectively, with an absolute isosurface value of $10^{-3}$ Å$^{-3}$.

## IV. CONCLUSION

Our HSE06 calculations identify thermodynamically stable neutral charge states and localized optical transitions in $X_{Si}V_C$ complexes (X = N, O, P, and S) in monolayer SiC. By assessing radiative and nonradiative decay, we find that $P_{Si}V_C$ is the leading SPE candidate among these complexes. It combines a large TDM (9.2 D) with a short radiative lifetime (18.2 ns) and negligible IC loss up to 800 K, supporting a near-unity quantum yield within the modeled radiative and IC channels. $N_{Si}V_C$ has the largest DW factor (52%), but multiphonon decay limits its predicted luminescence quantum yield to below 50%. $S_{Si}V_C$ also exhibits rapid radiative decay and negligible IC, but ISC through a lower-energy triplet could reduce its total quantum yield. Dopant-specific hyperfine tensors and localized spin densities provide experimentally testable signatures for identifying the spin-active $N_{Si}V_C$ and $P_{Si}V_C$ defects. Although previous screening studies of SPE candidates have often emphasized radiative metrics such as ZPL energy, transition dipole moment, and DW factor, nonradiative decay can quench emission and limit the quantum efficiency required for quantum applications. Our results establish a joint radiative–nonradiative screening strategy for rational SPE design and identify $P_{Si}V_C$ as a promising SPE in monolayer SiC.

**Acknowledgments**

This work was supported by the National Natural Science Foundation of China (Grant No. 12564036, 62674031), the Natural Science Foundation of Jiangxi Province (Grant No. 20252BAC250005), and the Science Challenge Project (Grant No. TZ2025013). The authors acknowledge the Hefei Advanced Computing Center for computational resources.

**AUTHOR DECLARATIONS**

**Conflict of Interest**

The authors declare no conflicts of interest.

**DATA AVAILABILITY**

The data supporting the findings of this study are available from the corresponding authors upon reasonable request.